\documentclass{eas}

\usepackage{graphicx}
\usepackage{amsmath}
\usepackage{amstext}
\usepackage[symbol]{footmisc}

\begin{document}

\vspace*{-2.6cm}
\hfill MZ-TH/05-18
\vspace*{2.1cm}
\TitreGlobal{Mass Profiles and Shapes of Cosmological Structures}



\vspace*{-1cm}
\title{Do we Observe Quantum Gravity Effects at Galactic Scales?}

\author{Reuter, M.}

\author{Weyer, H.}\address{Institute of Physics, University of Mainz, Staudinger Weg 7, 55099 Mainz, Germany}


%

\runningtitle{Quantum Gravity}

\setcounter{page}{1}


\index{Reuter, M.}

\index{Weyer, H.}



\begin{abstract}
  The nonperturbative renormalization group flow of Quantum Einstein
  Gravity (QEG) is reviewed. It is argued that there could be strong
  renormalization effects at large distances, in particular a scale dependent Newton
  constant, which mimic the presence of dark matter at galactic and
  cosmological scales.
\end{abstract}
\maketitle
\vspace*{-4.45cm}\hspace*{8cm}\footnote{Talk given by M.R.~at the 21st IAP meeting ``Mass Profiles and Shapes of Cosmological Structures'', Paris, July 4-9, 2005; to appear in the proceedings.}
\vspace*{3.974cm}
%




\section{Introduction}

By now it appears increasingly likely that Quantum Einstein Gravity
(QEG), the quantum field theory of gravity whose underlying degrees of
freedom are those of the spacetime metric, can be defined
nonperturbatively as a fundamental, ``asymptotically safe'' theory
(Lauscher 2002). By definition, its bare action is given by a
non--Gaussian renormalization group (RG) fixed point. In the framework
of the ``effective average action'' a suitable fixed point is known to
exist within certain approximations. They suggest that the fixed point
should also exist in the exact theory, implying its nonperturbative
renormalizability.

The general picture regarding the RG behavior of QEG as it has emerged
so far points towards a certain analogy between QEG and non--Abelian
Yang--Mills theories, Quantum Chromo--Dynamics (QCD) say. For example,
like the Yang--Mills coupling constant, the running Newton constant
$G=G(k)$ is an asymptotically free coupling, it vanishes in the
ultraviolet (UV), i.\,e.\ when the typical momentum scale $k$ becomes
large. In QCD the realm of asymptotic freedom is realized for momenta
$k$ larger than the mass scale $\Lambda_{\text{QCD}}$ which is induced
dynamically. In QEG the analogous role is played by the Planck mass
$m_{\text{Pl}}$. It delimits the asymptotic scaling region towards the
infrared (IR). For $k \gg m_{\text{Pl}}$ the RG flow is well described
by its linearization about the non--Gaussian fixed point. Both in QCD
and QEG simple local approximations (truncations) of the running
Wilsonian action (effective average action) are sufficient above
$\Lambda_{\text{QCD}}$ and $m_{\text{Pl}}$, respectively. However, as
the scale $k$ approaches $\Lambda_{\text{QCD}}$ or $m_{\text{Pl}}$
from above, many complicated, typically nonlocal terms are generated
in the effective action. In fact, in the IR, strong renormalization
effects are to be expected because gauge (diffeomorphism) invariance
leads to a massless excitation, the gluon (graviton), implying
potential IR divergences which the RG flow must cure in a dynamical
way. Because of the complexity of the corresponding flow equations it
is extremely difficult to explore the RG flow of QCD or QEG in the IR,
far below the UV scaling regime, by analytical methods. In QCD,
lattice results and phenomenology suggest that the nonperturbative IR
effects modify the classical Coulomb term by adding a confinement
potential to it which increases (linearly) with distance: $V (r) = -
a/r + \kappa \,r$.

The problem of the missing mass or ``dark matter'' is one of the most
puzzling mysteries of modern astrophysics. It is an intriguing idea
that the apparent mass discrepancy is not due to an unknown form of
matter but rather indicates that we are using the wrong theory of
gravity, Newton's law in the non--relativistic and General Relativity
in the relativistic case.  If one tries to explain the observed
non--Keplerian rotation curves of galaxies or clusters in terms of a
modified Newton law, a nonclassical term needs to be added to the
$1/r$-potential whose relative importance grows with distance. In
``MOND'', for instance, a point mass $M$ produces the potential $\phi
(r) = - G M / r + \sqrt{a_{0} \, G M \,} \, \ln (r)$ and it is
tempting to compare the $\ln(r)$-term to the qualitative similar
confinement potential in (quenched) QCD. It seems not unreasonable to
speculate that the ``confinement'' potential in gravity is a quantum
effect which results from the antiscreening character of quantum
gravity (Lauscher 2002) in very much the same way as this happens in
Yang--Mills theory. If so, the missing mass problem could get resolved
in a very elegant manner without the need of introducing dark matter
on an ad hoc basis. In (Reuter 2004a,b) this idea has been explored
within a semi--phenomenological analysis of the effective average
action of quantum gravity.

\section{RG running of the gravitational parameters}

\begin{figure}[h]
\centering \includegraphics[width=12cm]{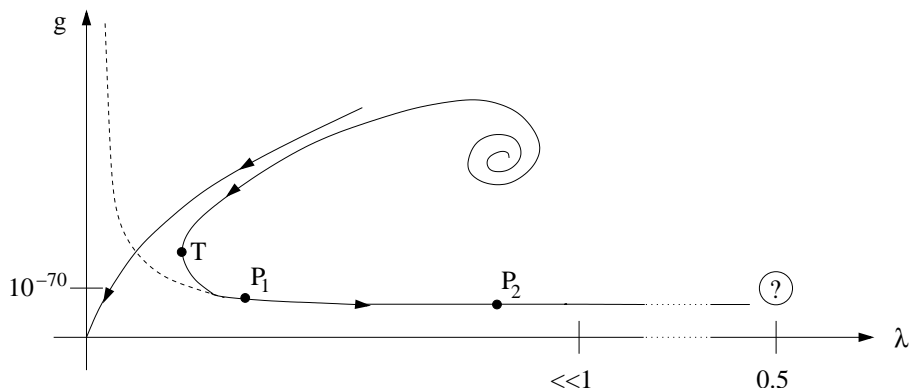}
\caption{Nature's Type IIIa trajectory and the separatrix.
The dashed line is a classical RG trajectory along which $G(k), \Lambda (k) = const$. (From (Reuter 2004b).)}
\label{fig1}
\end{figure}

The effective average action $\Gamma_{k} [g_{\mu \nu}]$ is a ``coarse
grained'' Wilsonian action functional which defines an effective field
theory of gravity at the variable mass scale $k$. Roughly speaking,
the solution to the effective Einstein equations $\delta \Gamma_{k} /
\delta g_{\mu \nu} =0$ yields the metric averaged over a spacetime
volume of linear extension $k^{-1}$. (From the technical point of view
$k$ is a IR cutoff introduced into the functional integral over the
microscopic metric in such a way that only quantum fluctuations of
wavelengths smaller than $k^{-1}$ are integrated out.) In a physical
situation with a typical scale $k$, the effective field equation
$\delta \Gamma_{k} / \delta g_{\mu \nu} =0$ ``knows'' about all
quantum effects relevant at this particular scale. For $k$ fixed, the
functional $\Gamma_{k}$ should be visualized as a point in the space
of all action functionals. When the RG effects are ``switched on'',
one obtains a curve in this space, the RG trajectory, which starts at
the bare action $S \equiv \Gamma_{k \to \infty}$ and ends at the
ordinary effective action $\Gamma \equiv \Gamma_{k \to 0}$. At the
exact level, $\Gamma_{k}$ contains all the infinitely many invariants
one can construct from $g_{\mu \nu}$, their $k$-dependent prefactors
having the interpretation of scale dependent gravitational coupling
constants. To become technically feasible most of the investigations
using the effective average action formalism employ the so--called
Einstein-Hilbert approximation which retains only Newton's constant
$G(k)$ and the cosmological constant $\Lambda(k)$ as running
parameters. If one introduces the dimensionless couplings $g(k) \equiv
k^2 G(k)$ and $\lambda(k) \equiv \Lambda(k)/k^2$ the RG equations
governing their scale dependence read $k \partial_k g = \beta_g(g,
\lambda)$, $k \partial_k \lambda = \beta_\lambda(g, \lambda)$ with
known beta--functions $\beta_g$ and $\beta_\lambda$. The RG flow on
the $g$-$\lambda$--plane displays two fixed points: a Gaussian fixed
point (GFP) at the origin, and the non-Gaussian fixed point (NGFP) at
$g_{*}>0$, $\lambda_{*}>0$ which is necessary for asymptotic safety.
The RG trajectories are classified as of Type Ia, IIa (separatrix),
and IIIa depending on whether, when $k$ is lowered, they run towards
negative, vanishing, and positive values of the cosmological constant,
respectively. In (Reuter 2004b) the very special trajectory which seems
realized in Nature has been identified and its parameters were
determined. This trajectory is of Type IIIa; see fig.\ \ref{fig1}.

For $k \rightarrow \infty$ it starts infinitesimally close to the
NGFP. Then, lowering $k$, the trajectory spirals about the NGFP and
approaches the ``separatrix'', the distinguished trajectory which ends
at the GFP. It runs almost parallel to the separatrix for a very long
``RG time''; only in the ``very last moment'' before reaching the GFP,
at the turning point T, it gets driven away towards larger values of
$\lambda$. In fig.\ \ref{fig1} the points P$_1$ and P$_2$ symbolize
the beginning and the end of the regime in which classical general
relativity is valid (``GR regime''). The classical regime starts soon after the
turning point T which is passed at the scale $k_{\rm T} \approx
10^{-30} m_{\mbox{\scriptsize Pl}}$.
   
In (Reuter 2004b) we argued that to the right of the point P$_2$ there
starts a regime of strong IR renormalization effects which might
become visible at astrophysical and cosmological length scales. In
fact, within the Einstein-Hilbert approximation, trajectories of Type
IIIa cannot be continued to the extreme IR ($k \rightarrow 0$). They
terminate at a non-zero value of $k$ as soon as the trajectory reaches
$\lambda = 1/2$. (Close to the question mark in fig.\ \ref{fig1}.)
Before it starts becoming invalid and has to be replaced by a more
precise treatment, the Einstein-Hilbert approximation suggests that
$G$ will increase, while $\Lambda$ decreases, as $\lambda \nearrow
1/2$.
   
The Type IIIa trajectory of QEG which Nature has selected is highly
special in the following sense. It is fine-tuned in such a way that it
gets {\it extremely} close to the GFP before ``turning left''. The
coordinates $g_{\rm T}$ and $\lambda_{\rm T}$ of the turning point are
both very small: $g_{\rm T} = \lambda_{\rm T} \approx 10^{-60}$. The
coupling $g$ decreases from $g(k) = 10^{-70}$ at a typical terrestrial
length scale of $k^{-1} = 1$ m to $g(k) = 10^{-92}$ at the solar
system scale of $k^{-1} = 1$ AU, and finally reaches $g(k) =
10^{-120}$ when $k$ equals the present Hubble constant $H_0$.
   
In fact, the Hubble parameter $k = H_0$ is approximately the scale
where the Einstein-Hilbert trajectory becomes unreliable. The
observations indicate that today the cosmological constant is of the
order $H_0^2$. Interpreting this value as the running $\Lambda(k)$ at
the scale $k = H_0$, the dimensionless $\lambda(k)$, at this scale, is
of the order unity: $\lambda(H_0) \equiv \Lambda(H_0)/H_0^2 =
\mathcal{O} (1)$.  So it is roughly near the present Hubble scale
where the IR effects should have grown large.
   
In principle it should be possible to work out the predictions of the
theory for cosmological scales by an ab initio calculation within QEG.
Unfortunately, because of the enormous technical complexity of the RG
equations, this has not been possible in practice yet. In this
situation one can adopt a phenomenological strategy, however. One
makes an ansatz for the RG trajectory which has the general features
discussed above, derives its consequences, and confronts them with the
observations. In this manner the observational data can be used in
order to learn something about the RG trajectory in the
nonperturbative regime which is inaccessible to an analytic treatment
for the time being. Using this strategy, the cosmological consequences
of a very simple scenario for the $k \to 0$ behavior has been worked
out; the assumption proposed in (Bonanno 2002) is that the IR effects
lead to the formation of a second NGFP into which the RG trajectory
gets attracted for $k \to 0$.  This hypothesis leads to a
phenomenologically viable late--time cosmology with a variety of
rather attractive features. It predicts an accelerated expansion of
the universe and explains, without any fine tuning, why the
corresponding matter and vacuum energy densities are approximately
equal.

\section{Galaxy rotation curves}

Given the encouraging results indicating that the IR effects are ``at
work'' in cosmology, by continuity, it seems plausible to suspect that
somewhere between solar system and cosmological scales they should
first become visible. In (Reuter 2004a,b) we therefore investigated
the idea that they are responsible for the observed non--Keplerian
galaxy rotation curves. The calculational scheme used there was a kind
of ``RG improvement'', the basic idea being that upon identifying the
scale $k$ with an appropriate geometric quantity comparatively simple
(local) truncations effectively mimic much more complicated (nonlocal)
terms in the effective action. Considering spherically symmetric,
static model galaxies only, the scale $k$ was taken to be the inverse
of the radial proper distance which boils down to $1 / r$ in leading
order. Since the regime of galactic scales turned out to lie outside
the domain of validity of the Einstein--Hilbert approximation the only
practical option was to make an ansatz for the RG trajectory $\big \{ G
(k), \Lambda (k), \cdots \big \}$ and to explore its observable
consequences. In particular a relationship between the $k$-dependence
of $G$ and the rotation curve $v (r)$ of the model galaxy has been
derived.

The idea was to start from the classical Einstein--Hilbert action and
to promote $G$ and $\Lambda$ to scalar fields:$S = \frac{1}{16 \pi} \,
\int \!\! \text{d}^{4} x~ \sqrt{-g\,} \big\{ R / G (x) - 2 \, \Lambda
(x) / G (x) \big \}$. Upon adding a matter contribution this action
implies the modified Einstein equation $G_{\mu \nu} = - \Lambda (x) \,
g_{\mu \nu} + 8 \pi \, G (x) \, \bigl( T_{\mu \nu} + \Delta T_{\mu
  \nu} \bigr)$ with $\Delta T_{\mu \nu} \equiv \frac{1}{8 \pi} \,
\bigl( D_{\mu} D_{\nu} - g_{\mu \nu} \, D^{2} \bigr) \, G^{-1}$.  In
(Reuter 2004a) we analyzed the weak field, slow--motion approximation
of this theory for a time--independent Newton constant $G = G
(\mathbf{x})$ and $\Lambda \equiv 0$. In this (modified) Newtonian
limit the equation of motion for massive test particles has the usual
form, $\ddot {\mathbf{x}} (t) = - \nabla \phi$, but the potential
$\phi$ obeys a modified Poisson equation:
\begin{align}
\nabla^{2} \phi = 4 \pi \,
\overline{G} \, \rho_{\text{eff}}
\quad \text{where }
\rho_{\text{eff}} \equiv
\rho + \bigl( 8 \pi \, \overline{G} \, \bigr)^{-1} \, \nabla^{2}
\mathcal{N}.
\end{align}
Here it is assumed that $T_{\mu \nu}$ describes
pressureless dust of density $\rho$ and that $G (\mathbf{x})$ does not
differ much from the constant $\overline{G}$. Setting $G (\mathbf{x})
\equiv \overline{G} \, \bigl[ 1 + \mathcal{N} (\mathbf{x}) \bigr]$ we
assumed that $\mathcal{N} (\mathbf{x}) \ll 1$. Apart from the rest
energy density $\rho$ of the ordinary (``baryonic'') matter, the
effective energy density $\rho_{\text{eff}}$ contains an additional
contribution $\bigl( 8 \pi \, \overline{G} \, \bigr)^{-1} \,
\nabla^{2} \mathcal{N} (\mathbf{x}) = \bigl( 8 \pi \, \overline{G}^{2}
\, \bigr)^{-1} \, \nabla^{2} G (\mathbf{x})$ due to the position
dependence of Newton's constant.  Since it acts as a source for $\phi$
on exactly the same footing as $\rho$ it mimics the presence of ``dark
matter''.

Up to this point the discussion applies to an arbitrary prescribed
position dependence of Newton's constant, not necessarily related to a
RG trajectory. In the case of spherical symmetry the natural choice of
the geometric cutoff is $k = \xi / r$ with $\xi$ a constant of order
unity. Hence we obtain the position dependent Newton constant $G
(\mathbf{x}) \equiv G (r)$ as $G (r) \equiv G ( k = \xi / r)$. Writing
again $G \equiv \overline{G} \, \left[ 1 + \mathcal{N} \, \right]$, $G
(k)$ should be such that $\mathcal{N} \ll 1$.

Let us make a simple model of a spherically symmetric ``galaxy''. For
an arbitrary density profile $\rho = \rho (r)$ the solution of the
modified Poisson equation reads
\begin{align}
\phi (r) = \int \limits_{}^{r} \!\! \text{d} r^{\prime}~
\frac{\overline{G} \, \mathcal{M} (r^{\prime})}{{r^{\prime}}^{2}}
+ \tfrac{1}{2} \, \mathcal{N} (r)
\label{48}
\end{align}
where $\mathcal{M} (r) \equiv 4 \pi \, \int_{0}^{r} \!\!
\text{d} r^{\prime}~ {r^{\prime}}^{2} \, \rho (r^{\prime})$ is the
mass of the ordinary matter contained in a ball of radius $r$. On
circular orbits test particles in the potential \eqref{48} have the
velocity $v^{2} (r) = r \, \phi^{\prime} (r)$ so that we obtain the
rotation curve
\begin{align}
v^{2} (r) = \frac{\overline{G} \, \mathcal{M} (r)}{r} 
+ \frac{1}{2} \, r \, \frac{\text{d}}{\text{d} r} \, \mathcal{N} (r).
\label{49}
\end{align}

We identify $\rho$ with the density of the ordinary luminous matter
and model the luminous core of the galaxy by a ball of radius $r_{0}$.
The mass of the ordinary matter contained in the core is $\mathcal{M}
(r_{0}) \equiv \mathcal{M}_{0}$, the ``bare'' total mass of the
galaxy. Since, by assumption, $\rho=0$ and hence $\mathcal{M} (r) =
\mathcal{M}_{0}$ for $r > r_{0}$, the potential outside the core, in
the halo, is $\phi (r) = - \overline{G} \, \mathcal{M}_{0} / r +
\mathcal{N} (r) / 2$.

As an example, let us adopt the power law $G (k) \propto k^{-q}$ with
$q>0$ which was motivated in (Reuter 2004a,b). We assume that this
$k$--dependence starts inside the core of the galaxy so that $G (r)
\propto r^{q}$ everywhere in the halo. For the modified Newtonian
limit to be realized, the position dependence of $G$ must be weak.
Therefore we shall tentatively assume that the exponent $q$ is very
small.  Expanding to first order in $q$ we obtain $\mathcal{N} (r) = q
\, \ln (r)$. In the halo, this leads to a logarithmic modification of
Newton's potential: $\phi (r) = - \overline{G} \, \mathcal{M}_{0} / r
+ \frac{q}{2} \, \ln (r)$.  The corresponding rotation curve is $v^{2}
(r) = \overline{G} \, \mathcal{M}_{0} / r + q/2$. At large distances
the velocity approaches a constant $v_{\infty} = \sqrt{q/2\,}$.
Obviously the rotation curve implied by the $k^{-q}$--trajectory does
indeed become flat at large distances -- very much like those we
observe in Nature.  Typical measured values of $v_{\infty}$ range from
$100$ to $300\,$km/sec, implying $q \approx 10^{-6}$ which is indeed
very small.  Including the core region, the complete rotation curve
reads $v^{2} (r) = \overline{G} \, \mathcal{M} (r) / r + q/2$. For a
realistic $\mathcal{M} (r)$ its $r$-dependence is in rough
qualitative agreement with the observations.

Our $v^{2} (r)$ is identical to the one obtained from standard
Newtonian gravity by postulating dark matter with a density
$\rho_{\text{DM}} \propto 1 / r^{2}$. We see that if $G (k) \propto
k^{-q}$ with $q \approx 10^{-6}$ no dark matter is needed. The
resulting position dependence of $G$ leads to an effective density
$\rho_{\text{eff}} = \rho + q / \bigl( 8 \pi \, \overline{G} \, r^{2}
\bigr)$ where the $1/r^{2}$--term, well known to be the source of a
logarithmic potential, is present as an automatic consequence of the
RG improved gravitational dynamics.

Thus it seems that if the observed non--Keplerian rotation curves are
due to a renormalization effect, the scale dependence of Newton's
constant should be roughly similar to $G (k) \propto k^{-q}$. Knowing
this, it will be the main challenge for future work to see whether a
corresponding RG trajectory is actually predicted by the flow
equations of QEG. For the time being an ab initio calculation of this
kind, while well--defined conceptually, is still considerably beyond
the state of the art as far as the technology of practical RG
calculations is concerned. In performing such calculations it might
help to rewrite the nonlocal terms generated during the flow in terms
of local field monomials by introducing extra fields besides the
metric. This is a standard procedure in the Wilsonian approach which
often allows for a simple local description of the effective IR
dynamics. It is tempting to speculate that the resulting local
effective field theory might be related to the generalized gravity
theory in (Papers I) which includes a Kalb--Ramond field; it is fully
relativistic and explains the galaxy and cluster data with remarkable
precision.

\end{document}